\documentclass[11pt]{article}
\usepackage{xspace}
\usepackage{graphicx}
\usepackage{amsmath}
\usepackage{amssymb}
\usepackage{color}

\definecolor{Red}{rgb}{1,0,0}
\definecolor{Green}{rgb}{0,1,0}
\definecolor{Blue}{rgb}{0,0,1}
\definecolor{Black}{rgb}{0,0,0}

\def\beq{\begin{equation}}
\def\eeq#1{\label{#1}\end{equation}}
\def\eeqn{\end{equation}}

\def\beqa{\begin{eqnarray}}
\def\eeqa#1{\label{#1}\end{eqnarray}}
\def\eeqan{\end{eqnarray}}

\let\bar=\overbar

\def\Dslash{\not{\hbox{\kern-4pt $D$}}}
\def\dslash{\not{\hbox{\kern-2pt $\del$}}}

\def\msb{{\bar{\ssstyle M \kern -1pt S}}}

\newcommand{\mecmueg}{\ensuremath{\mu\rightarrow e\gamma}\xspace}
\newcommand{\mecmec}{$\mu$--$e$ conversion\xspace}
\newcommand{\mecunit}[1]{\ensuremath{\,\textrm{#1}}\xspace}

\def\Title#1{\begin{center} {\Large {\bf #1} } \end{center}}

\begin{document}

\Title{Muon to electron conversion: The COMET and Mu2e experiments}

\bigskip\bigskip


\begin{raggedright}  

R. Phillip Litchfield\index{Litchfield, R.\,P.}, {\it University College, London}\\

\begin{center}\emph{Member of the COMET Collaboration.}\end{center}
\bigskip
\end{raggedright}

{\small
\begin{flushleft}
\emph{To appear in the proceedings of the Interplay between Particle and Astroparticle Physics workshop, 18 -- 22 August, 2014, held at Queen Mary University of London, UK.}
\end{flushleft}
}

\section{Charged lepton flavor violation}
\subsection{In the Standard Model}
The story of charged lepton flavour violation could be said to begin with the neutrino. As originally conceived, it is a massless particle that comes in three flavors, one for each charged lepton, as the other half of a weak isospin doublet. Since the Standard Model neutrino carries only weak charges, the weak eigenstates would be an appropriate basis for describing all neutrino phenomena and lepton flavor would be conserved in the theory.  However it is now firmly established that neutrinos do have masses, albeit very small ones, and the appropriate basis for a neutrino propagating in the vacuum need not be the same as the weak basis.  We find that the mixing between weak eigenstates and mass eigenstates is extremely large and, more recently, that all three mass states overlap with the weak states.  These deductions were made by the discovery and investigation of neutrino oscillations, which has become one of the most fruitful avenues of elementary physics research in the last decade.

\begin{figure}[!ht]
\centering
\begin{minipage}[c]{0.5\columnwidth}
\centering
\includegraphics[width=0.75\columnwidth]{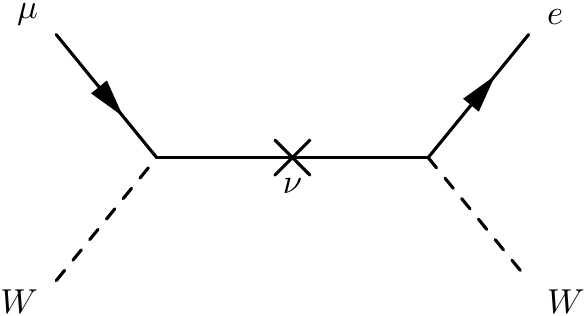}
\end{minipage}%
\begin{minipage}[c]{0.5\columnwidth}
\centering
\includegraphics[width=0.75\columnwidth]{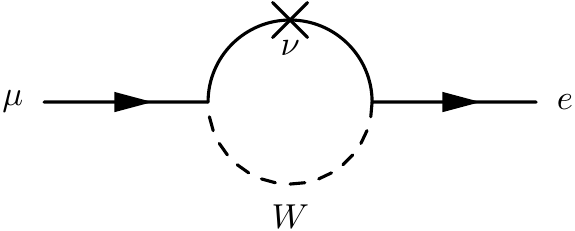}
\end{minipage}
\caption{(Left) The neutrino oscillation process; the `internal' neutrino typically travels several kilometres.  (Right) By connecting the initial and final $W$ the same process changes the flavour of a charged lepton.}
\label{fig:basic_clfv}
\end{figure}

But what then of the charged leptons?  If two different weak eigenstates of the neutrino can be coupled by their shared masses, it seems reasonable to imagine the same mixing would lead to a coupling of their charged counterparts.  We can quantify this na\"ive supposition, as shown in Figure~\ref{fig:basic_clfv}: If one takes the (tree level) diagram for neutrino oscillations, and imagine the two external $W$ boson lines are instead connected to each other, we have a loop diagram connecting leptons of different flavors.  Some additional interaction (most commonly electromagnetic radiation) must be involved to conserve energy and momentum, but there is no explicit reason why this process should not occur.  There are three kinematically allowed flavor variations, one involving a muon decaying to an electron, the other two being decays of the tau.  The muon to electron process is by far the easiest to study: the low mass of the muon limits the number of possible final states, and its long lifetime of the muon compared to the tau makes it considerably easier to use experimentally.  We focus here on muon conversion processes.

\begin{figure}[!ht]
\centering
\begin{minipage}[c]{0.5\columnwidth}
\centering
\includegraphics[width=0.75\columnwidth]{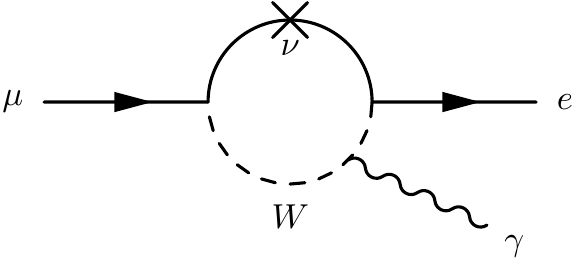}
\end{minipage}%
\begin{minipage}[c]{0.5\columnwidth}
\centering
\includegraphics[width=0.75\columnwidth]{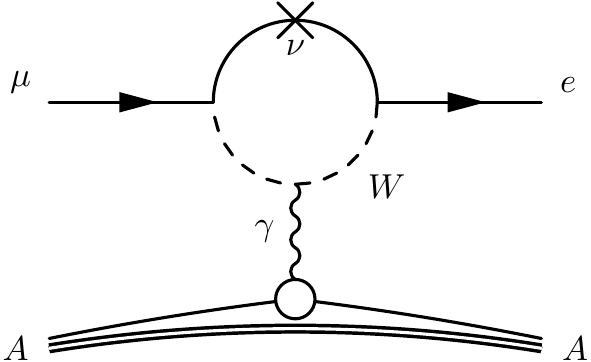}
\end{minipage}
\caption{(Left) The leading-order Standard Model $\mu \rightarrow e\gamma$ process, with the photon arbitrarily attached to the $W$ boson  line.  (Right) A very similar diagram contributes to the $\mu$--$e$ conversion process. }
\label{fig:sm_clfv}
\end{figure}

It turns out that although not forbidden, these processes are extremely rare in the Standard Model because of the large difference in mass of the particles in the loop.  The neutrino propagator suppresses high values of $q^2$ around the loop, while the $W$ propagator suppresses low values.  The overall effect is that the process is heavily suppressed relative to the dominant decay modes.  For the simplest neutrinoless decay of a free muon (into an electron plus photon, see Fig.~\ref{fig:sm_clfv}), the Standard Model branching ratio is 
\begin{equation}
\frac{\Gamma(\mu\rightarrow e\gamma)}{\Gamma(\mu\rightarrow e\nu\nu)}\propto\left|\sum_i \frac{m_i^2}{m_W^2}U^*_{\mu i}U_{e i}\right|^2 \sim 10^{-54}\textrm{,}
\end{equation}
where the $m_i$ are the neutrino masses, and $m_W$ is the $W$ boson mass.  This is many orders of magnitude below the sensitivity of any currently conceived experiment, and so a flavor symmetry of charged leptons appears to exist in the Standard Model.

The muon to electron ($\mu$--$e$) conversion process is very similar, as emphasised by Fig.~\ref{fig:sm_clfv}.  The photon is absorbed coherently by the nucleus, which remains in the same state but recoils slightly, providing the freedom to conserve energy and momentum.  Because of the large nuclear mass, the emitted electron carries almost all the energy of the muon decay and is practically mono-energetic.

\subsection{In new physics scenarios}
\begin{figure}[!ht]
\centering
\begin{minipage}[c]{0.5\columnwidth}
\centering
\includegraphics[width=0.75\columnwidth]{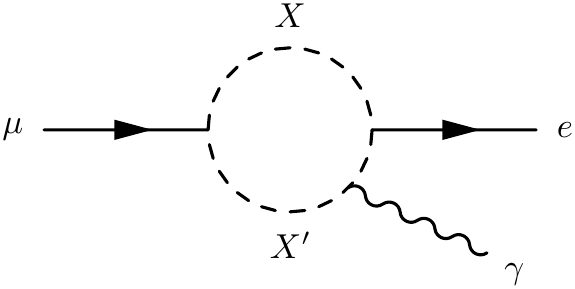}
\end{minipage}%
\begin{minipage}[c]{0.5\columnwidth}
\centering
\includegraphics[width=0.75\columnwidth]{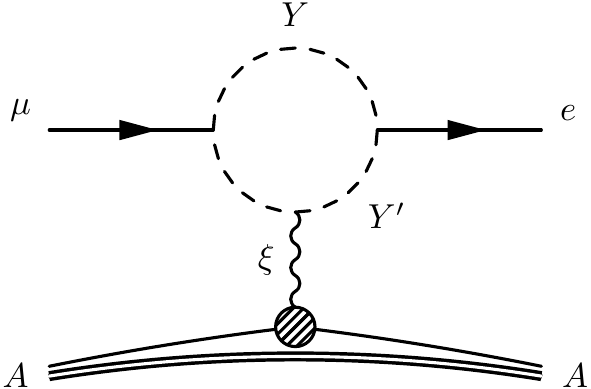}
\end{minipage}
\caption{New physics analogues of Figure~\ref{fig:sm_clfv}.  The nature of particles $X$, $X^\prime$, $Y$, $Y^\prime$ and $\xi$ are unspecified, and depend on the new physics scenario.}
\label{fig:new_clfv}
\end{figure}

As extensions to the Standard Model, many new physics scenarios provide fields that can take the place of the $W$ and neutrino in the internal loop of a charged lepton flavor-changing process. The new physics analogues are shown schematically in Fig.~\ref{fig:new_clfv}.  As with the Weak interaction, there is no reason to expect new physics to respect lepton generation number, so in general these also cause flavor violation.  Moreover, the suppression from extreme mass disparity that occurs for the Standard Model is not generally a feature of most new physics loops, so they quite naturally predict much higher branching ratios.  Thus searches for charged lepton flavor violation have good sensitivity to a very broad range of models.

\begin{figure}[!ht]
\centering
\begin{minipage}[b]{0.5\columnwidth}
\centering
\includegraphics[width=0.75\columnwidth]{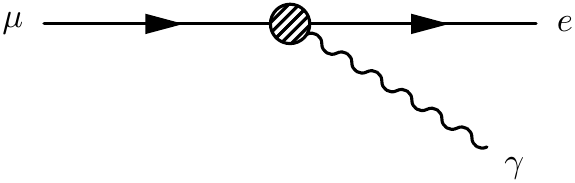}
\includegraphics[width=0.75\columnwidth]{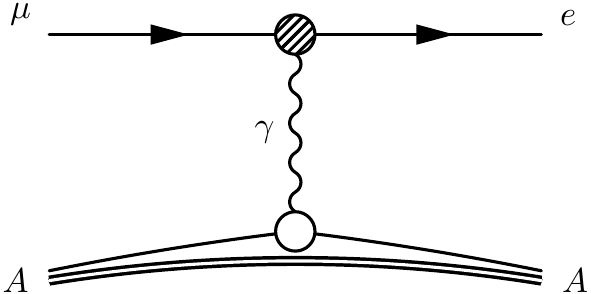}
\vspace{0.2cm}\\
$$\mathcal{L}_d \sim \frac{1}{\Lambda^2}\, m_\mu \, \overline{\mu}_L \sigma_{\mu\nu}e_R \, F^{\mu\nu}$$
\centering
\end{minipage}%
\begin{minipage}[b]{0.5\columnwidth}
\centering
\includegraphics[width=0.75\columnwidth]{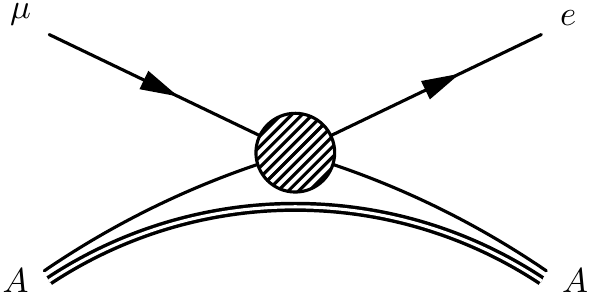}
\vspace{0.2cm}\\
$$\mathcal{L}_4 \sim \frac{1}{\Lambda^2}\, \overline{\mu}_L \gamma_{\mu}e_L \,
\left(\overline{u}_L \gamma^{\mu} u_L + \overline{d}_L \gamma^{\mu} d_L\right)$$
\end{minipage}
\caption{Effective field interactions for flavour-violating muon decays. (Left) A dipole operator can induce $\mu\rightarrow e\gamma$ and $\mu$--$e$ conversion.  (Right) A four-fermion interaction can only induce $\mu$--$e$ conversion.}
\label{fig:eff_clfv}
\end{figure}

The new physics mediators of these processes can be assumed, essentially by definition, to be substantially heavier than the muon.  Therefore we can reduce these diagrams to effective field theories, with a mass scale $\Lambda$.  If the new physics occurs \emph{only} in the loop, the photon analogue (labelled $\xi$) is just a photon and the effective field theory includes a new lepton-photon vertex that changes lepton flavour.  Conversely, if the exchanged $\xi$ is a heavy new physics mediator the effective field theory will include a four-fermion point interaction. These alternatives are shown in Fig~\ref{fig:eff_clfv}.  It should be emphasised that while we have used analogues of the virtual $W$--$\nu$ loop for illustrating the possible effective field theory interactions, the resulting couplings encompass a range of other possible processes, particularly in the case of the four-fermion interaction.  

\begin{figure}[!ht]
\centering
\raisebox{0.44\columnwidth}{\includegraphics[width=0.15\columnwidth]{mueg_eff.pdf}}
\includegraphics[width=0.5\columnwidth]{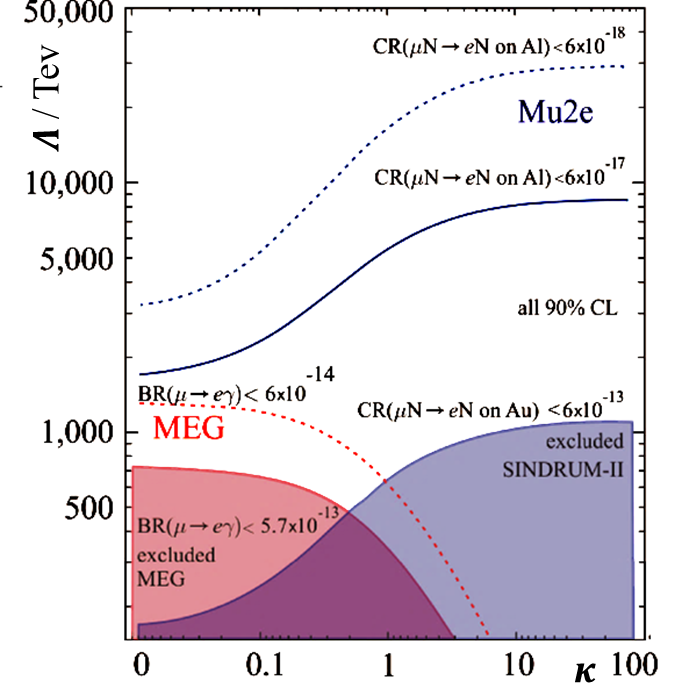}
\raisebox{0.43\columnwidth}{\includegraphics[width=0.15\columnwidth]{mec_ff_eff.pdf}}
\caption{Current limits (filled) and future sensitivity (lines) to Charged Lepton Flavor Violating processes as a function of the parameter $\kappa$. Figure adapted from one provided by Mu2e; COMET has the same dependence and sensitivities are similar.}
\label{fig:kappa_dep}
\end{figure}

A general Lagrangian including both terms can be parametrised \cite{deGouva:2009zz, Kuno:1999jp} with a relative strength parameter, $\kappa$:
\begin{equation}
\mathcal{L}_{\mu e} \sim \frac{1}{\Lambda^2}\left[\frac{1}{1+\kappa}\, m_\mu \,\overline{\mu}_L \sigma_{\mu\nu}e_R \, F^{\mu\nu} + \frac{\kappa}{1+\kappa}\,\overline{\mu}_L \gamma_{\mu}e_L  \, \left(\overline{u}_L \gamma^{\mu} u_L + \overline{d}_L \gamma^{\mu} d_L\right)\right]\textrm{.}
\end{equation} 
This conveniently incorporates the the physics for from both \mecmueg and \mecmec.  The parameter $\kappa$ (or a range of allowable values) is characteristic of a new physics model, so the discovery potential of each measurement depends on the model, as shown in Fig~\ref{fig:kappa_dep}.  In the event of a positive signal, comparison between the two channels can be used to measure $\kappa$ and narrow down the possible models.  For a given value of $\Lambda$, a small value of $\kappa$ means the dipole interaction is more important, which favours the \mecmueg channel. Conversely, for large value of $\kappa$ the \mecmec conversion channel is more sensitive.  In particular, if a new physics model gives rise to a $\kappa$ that is very large, the sensitivity of the \mecmueg can be arbitrarily suppressed.  Nevertheless, this channel has historically provided the best sensitivity for many models, with the most recent results coming from the MEG experiment at PSI\cite{Adam:2013mnn}.  Because these kinds of experiments have relied on a coincidence between electron and photon signals, increasing the muon intensity raises the combinatoric background, making improvements beyond the next MEG upgrade infeasible with current technology.  Detection in \mecmec does not require a coincidence and this restriction does not apply, so a new generation of \mecmec experiments, COMET\cite{cometweb} and Mu2e\cite{mu2eweb}, are in development to take advantage of new high intensity pulsed muon beams being constructed at J-PARC and Fermilab, respectively. 

\section[Basics of a (modern) $\mu$--$e$ conversion experiment] {Basics of a (modern) 
$\boldsymbol{\mu}$--$\boldsymbol{e}$ conversion experiment}{
The new generation of \mecmec experiments utilise a low-energy negative muon beam that is allowed to stop in a suitable target. Stopped muons are captured by atoms in the target and rapidly cascade down to the muonic 1s atomic orbital.  The conversion interaction is coherent on the nucleus, which absorbs very little energy as it recoils, thus the electron is effectively mono-energetic at $105\mecunit{MeV} (\simeq m^-_\mu-B^\mu_{1s} )$.  

\subsection{Backgrounds}
The normal decay of a orbiting muon is nuclear muon capture ($\mu N(Z)\rightarrow \nu_\mu N(Z-1)$), which produces no background.  Alternatively the muon may decay in orbit (DIO: $\mu^-\rightarrow \nu_\mu \overline{\nu}_e \ e^-$), which emits an electron.  For a free muon the electron can have an energy of no more than $\frac{1}{2}m_\mu c^2$, but bound in a muonic atom there is a small tail, of order 1 in billions, of more energetic electrons up to the conversion energy, where the neutrinos are at rest and the electron recoils against the nucleus. Although relatively rare, this is the most important background for the conversion process, and in order to mitigate it the experiments require good energy resolution.

The other way a decay electron can gain enough energy to be a background at 105 MeV is to be boosted in the lab frame by a muon decaying in flight, outside the target.  The boost must be large enough to replace the energy lost to the neutrinos, so this background can be suppressed by designing the muon beam transport to have low transmission for muons over $\sim\!60\mecunit{MeV/$c$}$. This can also be reduced by tracking detectors that can identify electrons coming from the target.

A final important background source is charged particles produced directly from the muon production target.  Unlike muon decays which are delayed by the decay lifetime of the muon in the target material, these prompt backgrounds will arrive close in time to the primary beam.  This is why pulsed beams are used in the new generation of experiments: by adopting a delayed signal window the prompt backgrounds can be easily eliminated.  This creates a tension however: the muon conversion process is coherent on the nuclear target and enhanced by a smaller 1s orbital, favouring the use of heavier nuclei as targets.  But as a result, heavier target nuclei reduce the muon lifetime, making separation from the prompt background more difficult.  The previous experiment, SINDRUM-II \cite{Bertl:2006up}, used a continuous beam and a gold target, but to get better rejection of the prompt background, both COMET and Mu2e will start with aluminium ($Z=13$) targets, because the lifetime ($\tau_\mu = 880\mecunit{ns}$) is better matched to their pulsed beam structure.   

For the use of a delayed window to be effective the, extinction of the beam (ratio of the number of particles outside of a filled accelerator bucket, compared to in a filled bucket) must be very low.  For COMET this was recently tested at the J-PARC Main Ring abort line. The required level is below $10^{-9}$; Fig.~\ref{fig:extinction} shows that this was comfortably exceeded, and that extinction down to $10^{-12}$ was possible, depending on the voltage of the applied RF field.
\begin{figure}[!ht]
\centering
\includegraphics[width=0.70\columnwidth]{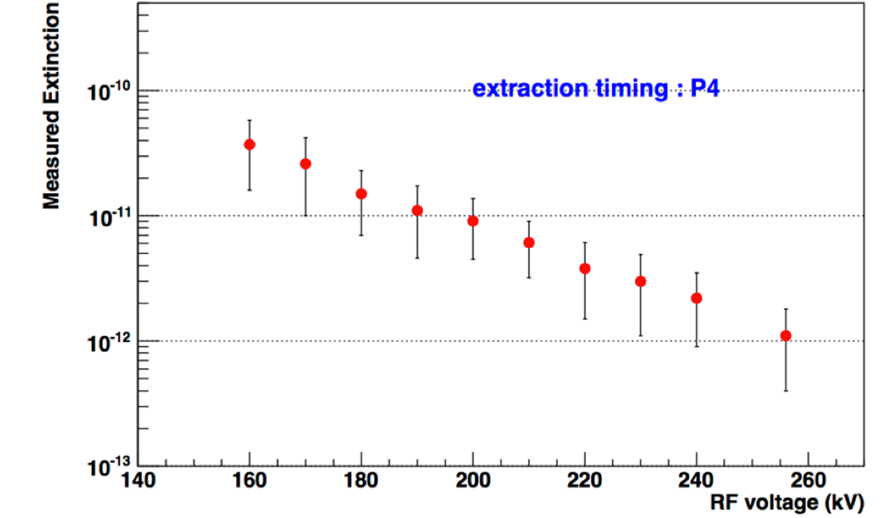}
\caption{Results of extinction tests at the J-PARC Main Ring abort line in 2014. The requirement of $10^{-9}$ was exceeded even with the lowest attempted RF voltage.}
\label{fig:extinction}
\end{figure}

\subsection{General features of experimental design}
If background are controlled, the main constraint on sensitivity is the need for a large number of low energy muons.  Both experiments will use dedicated high-power beam lines that can provide pulses of protons at around $8\mecunit{GeV}$. A resonant slow extraction scheme is used to take a small fraction of the circulating pulses in each cycle.  The extracted protons are directed onto a low-Z production target, which sits inside a powerful superconducting solenoid.  The solenoid field is shaped to provide a large solid angle angle of acceptance in the backward direction, to collect low-energy pions and funnel them into the secondary beam transport.  The high beam currents involved mean that many neutrons are produced, which necessitates substantial shielding, and consideration given to heating of the solenoid inner bore.

The transport line in which pions decay to produce muons is one of the most characteristic features of these new experiments.  Both use curved solenoids that keep the low energy muons on helical trajectories while deflecting them through $90^\circ$ in the horizontal plane.  The curved transport line eliminates the direct line of sight from the production target to the muon stopping target, allowing the positioning of shielding to reduce the neutron background.  The curved solenoids also have another, more specific, purpose.  In deflecting the helical trajectories a net drift is also created in the vertical plane, which depends on the charge sign and momentum of the particles in the beam. This removes positive and high-energy muons and other particles that contribute to the backgrounds, and the effect can be enhanced with a collimator.  The exact arrangement of these components is the most obvious difference between the two experiments.

\section{Mu2e}
The positioning of the curved transport lines determines the experiments' overall shape.  Mu2e uses two $90^\circ$ transport sections arranged in an `S' shape, as shown in Fig.~\ref{fig:mu2e}, so that the largest beam dispersion is at the midpoint of the transport, the beam being returned to horizontal at the end of the second bend.  A vertically offset collimator located at the midpoint is used to remove the unwanted positive and higher-momentum components.  This collimator can be rotated for background studies. 
\begin{figure}[!ht]
\centering
\includegraphics[width=0.9\columnwidth]{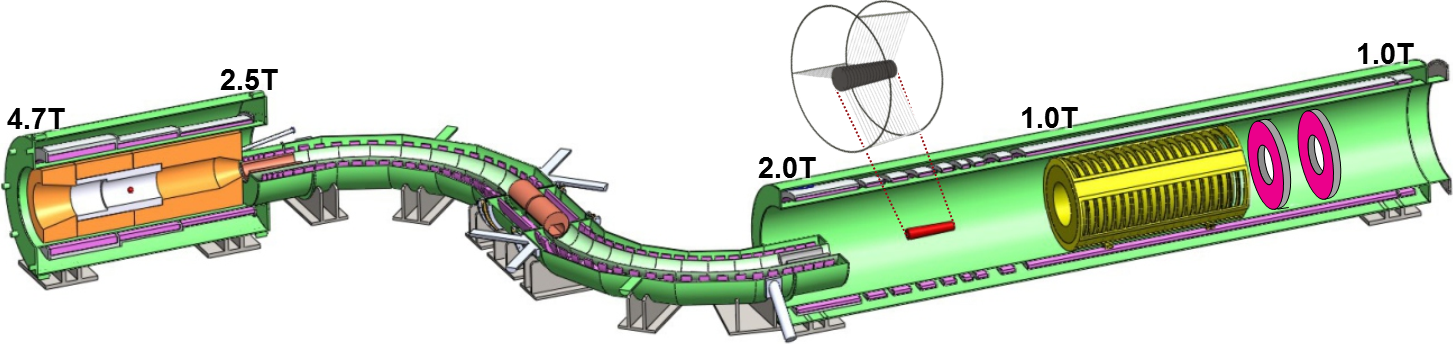}
\caption{Layout of the Mu2e experiment.  The experiment is shown with the production target and solenoid on the left and the detector solenoid on the right.}
\label{fig:mu2e}
\end{figure}

The capture solenoid of Mu2e uses a gradient field that has a maximum of 4.7\mecunit{T} at the forward end down to 2.5\mecunit{T} at the backward end.  This design reflects lower momentum particles from the forward direction to the backward direction, in order to increase the yield.  

At the other end of the transport line there is a long solenoid that holds the muon capture target and detectors.  The target consists of 17 Aluminium foils of $0.2\mecunit{mm}$ thickness and sits on the solenoid axis.  Two detectors sit beyond the target: Closer to the target is an 80-plane Straw Tube Tracker and beyond this is a double ring Electromagnetic Calorimeter.  For the energy regime of a muon decay, the combination of momentum and energy measurements allows excellent particle identification in off-line analysis.
 
Because of the beam passing near the axis of the solenoid, and the substantially larger numbers of DIO electrons in the inner regions, the detectors have annular geometry and do not cover the central part of the solenoid.  The radius of the helical paths increases with energy, and the detectors are designed to sample only the higher energy tracks, as shown in Fig.~\ref{fig:mu2e_radius}. Because the paths of even high energy particles will pass repeatedly into the uninstrumented inner region, there will be significant gaps in the track. This is why the Straw Tube Tracker has so many planes and there are two rings of calorimeter.
\begin{figure}[!ht]
\centering
\includegraphics[width=0.47\columnwidth]{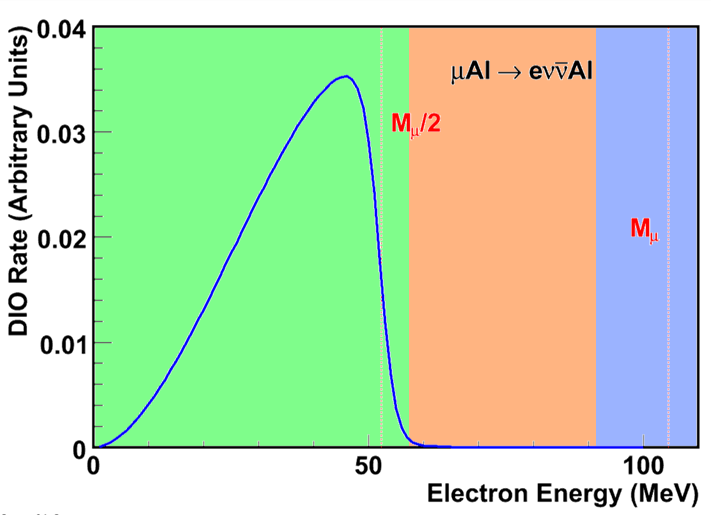} \quad
\includegraphics[width=0.33\columnwidth]{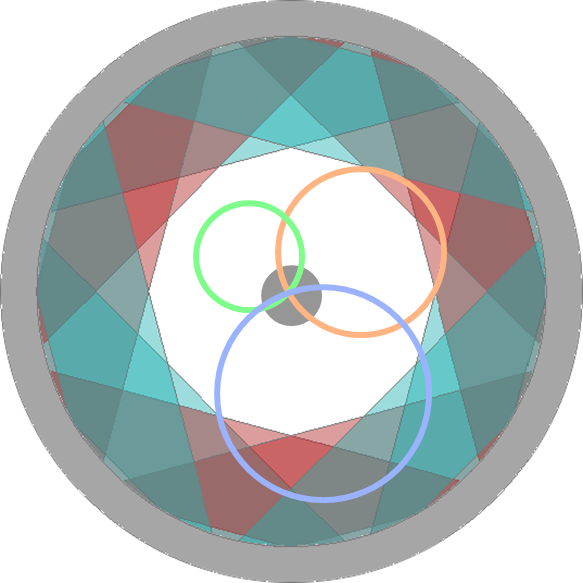}
\caption{Track sampling of the Mu2e straw chambers.  The frontal view of the Straw Tube Tracker is shown on the right, and the muon DIO spectrum on the left. The coloured circles in correspond to typical helix radii of electrons in the corresponding energy region.}
\label{fig:mu2e_radius}
\end{figure}

\section{COMET}
Although in many respects COMET and Mu2e are very similar, there are some significant differences in their design.  COMET is design to run in two phases, with different geometries. The layout of the two phases is shown in Fig.~\ref{fig:comet_layout}.  Phase-I of the experiment uses a $90^\circ$ muon transport line which filters out less background. This is sufficient to make an intermediate-sensitivity measurement on a shorter time-scale.  Perhaps more importantly, it can also be used to characterise the backward pion yield from the production target and improve the simulations used for use in Phase-II.  The Phase-II experiment uses a longer transport line, which curves round $180^\circ$ in a `C' shape. This creates a larger momentum dispersion in the vertical plane, improving the rejection of wrong-momentum muons.  Unlike Mu2e there is no `reverse turn' to level out the beam. COMET's beam transport instead uses a compensating dipole field to keep the desired lower-momentum negative muons vertically centred in the transport line.     
\begin{figure}[!ht]
\centering
\includegraphics[width=0.9\columnwidth]{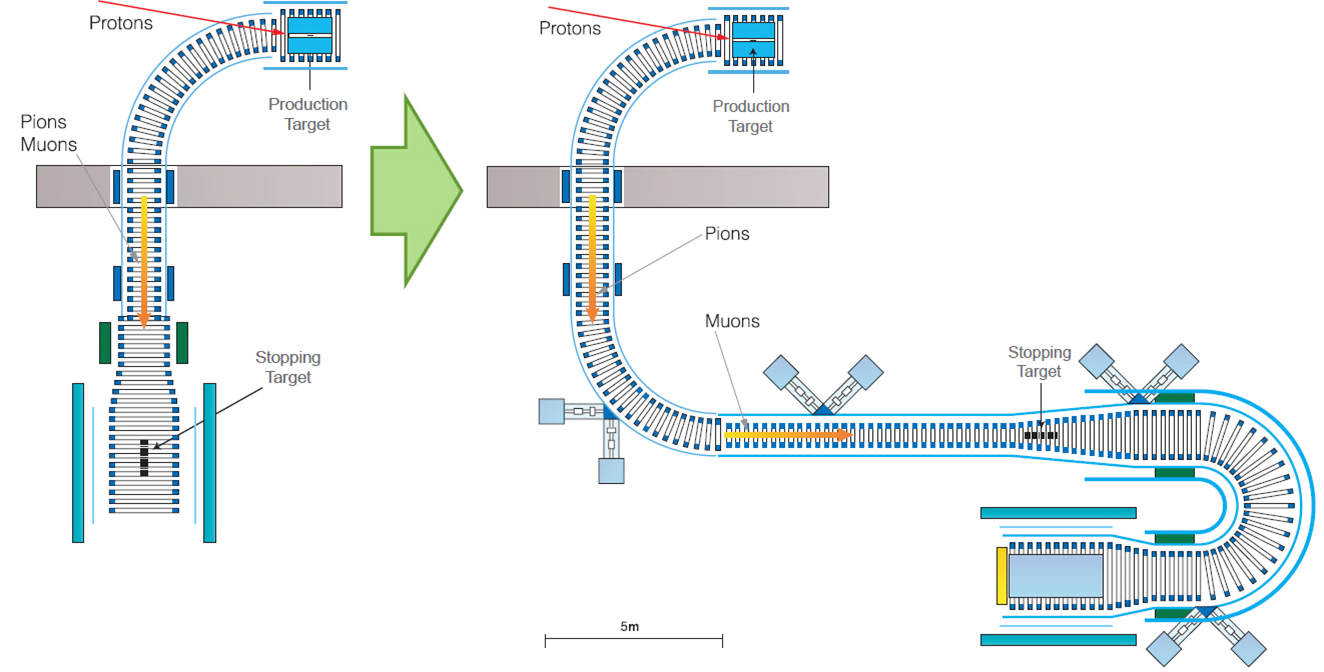}
\caption{The layout of COMET Phase-I (Left) and Phase-II (Right). The upper part of the diagram shows the pion capture solenoid, and the first $90^\circ$ transport passing through the shield wall; these elements are common to both phases.  In Phase-I these lead direct to a target surrounded by a detector, while in Phase-II there is a longer transport before the target, and---before the detectors---another curved solenoid that acts as an electron spectrometer.}
\label{fig:comet_layout}
\end{figure}

The pion capture solenoid in COMET has a similar peak field strength of 5\mecunit{T}, but unlike Mu2e it peaks at the target location, rather than at the end of the solenoid.  As a result there is no magnetic reflection to capture forward going pions; this configuration instead has wider solid angle of acceptance, for pions emitted at a larger angle from the solenoid axis.

\subsection{Phase-I setup}
In Phase-I, the beam will characterised by prototype sections of the Phase-II detectors (discussed below).  These measurements should be made before the final design of Phase-II so that it can be optimised with a fully tuned simulation of the backward-scattered pions.  In addition the Phase-I set-up naturally provides a better environment for beam measurements as the shorter transport length lets us observe shorter-lived components of the secondary beam.  

Due to smaller acceptance and the large beam background at their location, these detectors are not intended for physics measurements in Phase-I, so another detector is included: a cylindrical drift chamber that sits inside the solenoid, surrounding  muon-stopping target. Like the Mu2e detector its inner radius is chosen to avoid backgrounds from low-energy electrons and beam particles close to the beam axis. The wires of the drift chamber are nearly parallel to the solenoid axis (but use a small stereo angle for measurement of the longitudinal co-ordinate). Hodoscope rings mounted on the inner wall of the drift chamber at its upstream and downstream ends will provide an event  trigger.

\subsection{Phase-II setup}
In Phase-II, the first major upgrade is the longer `C' shaped muon transport.  This reduces the background rates further by providing a narrower band pass on particle momentum.  The other major change is that instead of sitting inside or directly in front of the detectors (as in Phase-I or Mu2e), Phase-II of COMET will place the stopping target in the throat of a another `C' shaped curved solenoid, this one designed to accept electrons with momentum of around 100\mecunit{MeV}, thus acting as a spectrometer and performing a signal pre-selection.  Beyond this electron spectrometer is a Straw Tube Tracker for momentum measurement and Electromagnetic Calorimeter.  As in the Mu2e design, combining energy and momentum measurements gives excellent discrimination between electrons and other particle types.  

Because the combination of muon transport tuned for $p\lesssim60\mecunit{MeV/$c$}$ and electron spectrometer tuned for $p\sim100\mecunit{MeV/$c$}$ eliminates much of the background, and means that the Comet Phase-II detector will be able to instrument the entire area of the solenoid without any minimum radius.  Because of this it needs only 5 (2-D) tracking stations to perform track reconstruction with the necessary momentum resolution. Similarly, in this design only a single calorimeter plane is required.

\section{Timelines and outlook}
Both Mu2e and COMET occupy sizeable new facilities at their hosting laboratories, and both are already under construction. COMET will be housed in an extension to J-PARC's Hadron physics Experimental Hall, and civil construction is progressing rapidly.  For COMET the goal is to start Phase-I in 2016, and run for around 100 days before beginning construction of Phase-II, which in turn would start in 2019 and run for a longer period.  Mu2e will start later, with a goal of early 2020, as they will use the same location as the new $g$$-$$2$ experiment at Fermilab, which is currently being installed.  

The most recent search for \mecmec by SINDRUM-II was in 2004, and had a branching ratio sensitivity of $\mathcal{O}(10^{-12})$. In the intervening time much improvement has been made in the similar \mecmueg channel, but further advances are technologically difficult.  Using new techniques in the \mecmec channel, both COMET and Mu2e expect to be able to improve on the SINDRUM-II limit by around four orders of magnitude in the next $5\sim10$ years.  Because the standard electroweak process is so heavily suppressed by the `accidental' smallness of neutrino masses, any positive signal is an indication of new physics, and signals can be expected from almost any new physics coupling to the lepton sector.  If a positive signal is seen further investigation of muon decay channels can also reduce the space of possible models.



\bigskip
\section*{Acknowledgments}
I am grateful to Jim Miller and John Quirk of the Mu2e collaboration for providing information and materials relating to their experiment.  My work on the COMET experiment is supported by UCL and STFC (UK) and I additionally thank NSFC (China), JSPS and MEXT (Japan) and MES (Russia) and other funding agencies for support of the experiment.

%
%

%
%
%
%
 
\end{document}